\documentclass[amsmath,amssymb,amsfonts,twocolumn,english,showpacs,superscriptaddress,prb]{revtex4-1}
\pdfoutput=1

\usepackage[T1]{fontenc}
\usepackage[latin9]{inputenc}
\usepackage{times}
\usepackage{graphicx}
\usepackage{color}
\usepackage{mathrsfs}
\usepackage{float}
\usepackage{indentfirst}
\usepackage{babel}
\usepackage{wasysym}
\usepackage{tikz}
\usepackage{amsthm}
\usetikzlibrary{calc}
\usepackage{verbatim}
\usepackage{tabularx}
\usepackage{tabulary}
\usepackage{mwe}
\usepackage[normalem]{ulem}

\usepackage{booktabs}
\usepackage{array, multirow}
\usepackage{subfigure}

\usepackage[colorlinks,bookmarks=false,citecolor=blue,linkcolor=blue,urlcolor=blue]{hyperref}
\setcitestyle{numbers,square}
\usepackage{array}
\usepackage{gensymb}

\graphicspath{{.}}

\begin{document}

\title{Probing viscous regimes of spin transport with local magnetometry}

\author{Jun Ho Son}
\affiliation{Department of Physics, Cornell University, Ithaca, New York 14853, USA}

\begin{abstract}
It is now well-established, both theoretically and experimentally, that charge transport of metals can be in a hydrodynamic regime in which frequent electron-electron collisions play a significant role. Meanwhile, recent experiments  have demonstrated that it is possible to inject spin currents into magnetic insulator films and explore the DC transport properties of spins. Inspired by these developments, we investigate the effect of viscosity, which naturally arises in the hydrodynamic regime, on DC spin transport. We show that viscosity gives rise to a sharp peak in the spatial profile of the out-of-plane stray magnetic field near the spin current injector. We propose that local magnetometers such as SQUIDs and nitrogen-vacancy centers can detect this viscosity-induced structure in the stray magnetic field. We also discuss the relevance of our results to yittrium iron garnet, a ferromagnetic insulator, and to Kagome spin liquids. 
\end{abstract}

\maketitle

\section{Introduction}

 Recent experiments \cite{spint1,spint2} have demonstrated that spin currents can be injected into magnetic insulator thin films by interfacing them with metals exhibiting a strong spin Hall effect. In this setup, an electric current through the metal generates a transverse spin current via the spin Hall effect; this spin current can propagate through the bulk of the magnetic insulator if it supports long-range, low-energy excitations carrying spin quantum numbers. Although spin is not strictly conserved due to dipolar interactions, spin-orbit couplings, and etc., these experiments show that in certain magnetic materials, one can observe coherent spin transport signals up to several micrometers away from the spin current sources. These developments open a new avenue for studying the low-energy physics of quantum magnets with ideas and methods inspired by charge transport, some of which have been explored theoretically in the context of ordered magnets \cite{Takei1,Takei2} and quantum spin liquids \cite{Chen2013,Chatterjee2015,Zhuang2021}. 

 In parallel, it has long been posited theoretically that some metals can enter a hydrodynamic regime in which momentum-conserving collisions between electrons dominate over momentum-relaxing ones \cite{Gurzhi, Kivelson2011}. In such a metal, viscosity plays an important role in charge transport, drastically altering the spatial profile of electric currents and potentials in transport experiments \cite{Gurzhi, nonlocal1,nonlocal2,nonlocal3,nonlocal4,nonlocal5,Haoyu1,graphene1,graphene2,Superballistic,visual}. A more contemporary series of experiments has identified the hydrodynamic regime in clean 2D GaAs electron gas \cite{dejong1995}, charge-doped graphene \cite{nonlocal5, graphene2016}, and many more. These discoveries have established hydrodynamics -- long regarded as a theoretical framework for strongly interacting quantum systems \cite{Kadanoff,Damle1997} -- as a powerful tool for studying quantum phases of matter.

  More recently, theoretical proposals \cite{spinhydro1,spinhydro3} and an experimental discovery of hydrodynamic sound modes through magnetic noise measurements \cite{spinhydro2} suggest that spin transport can also be in a hydrodynamic regime. Together with the aforementioned experimental advances, these works motivate a natural question: What would be the experimentally accessible signature of the hydrodynamic DC spin transport? 
 
 Identifying such signatures in quantum magnets will provide new opportunities to explore exotic phases of matter governed by \textit{many-body dynamics} of spin-carrying quasiparticles, such as magnons and fractionalized spinons. In particular, experimental techniques historically used to probe magnetic materials, such as susceptibility measurements and neutron scattering, have been powerful enough to study conventional ordered magnets but have fallen short of properly characterizing more exotic magnetic phases featuring non-trivial topology and long-range entanglement-- most notably quantum spin liquids \cite{spinliquidreview}. Developing a probe of hydrodynamic spin transport will endow us with a potent method of studying these phases of matter.  
 
Here, we combine the spin current injection setup previously employed in the experiments in Ref.~\onlinecite{spint1} and \onlinecite{spint2} with local magnetometers such as superconducting quantum interference devices (SQUIDs) and nitrogen-vacancy (NV) centers to address this question; see Fig.~\ref{fig:fig1}(a) for the illustration. To see how hydrodynamic effects show up in this setup, we theoretically study the spatial profiles of the spin chemical potential \cite{chemical1,chemical2} and the stray magnetic fields across different transport regimes. Specifically, we consider the following three spin transport regimes: $(i)$ A diffusive regime, governed by standard diffusion equations commonly used to simulate spin transport experiments \cite{chemical1} $(ii)$ a 3D hydrodynamic regime described by linearized Navier-Stokes equations $(iii)$ a 2D hydrodynamic regime captured by 2D linearized Navier-stokes equations augmented with out-of-plane diffusion of momentum and spins. Our study analysis reveals  that in both hydrodynamic regimes, due to viscosity, the out-of-plane component of the stray magnetic field $B_{\perp}$ features a sharp peak near the spin current injector.  The height of this peak is sensitive to the vertical distance from the sample and is detectable via NV centers or SQUIDs. Notably, this feature is absent in the purely diffusive regime.

 The results on 3D hydrodynamic regime are expected to be relevant to the magnon transport in the isotropic ferromagnetic insulator such as yittrium iron garnet (YIG); YIG can be grown ultra-clean so that magnon mean free paths are set by momentum-conserving magnon-magnon collsions at sufficiently high temperature \cite{spinhydro1}. Meanwhile, the 2D hydrodynamic regime may be realized in quantum spin liquid candidates on Kagome lattices, such as Herbertsmithite. Kagome spin liquids have previously been proposed to be Dirac spin liquids \cite{Dirac} in which spinon Umklapp scattering is suppressed due to the Dirac-like dispersion of spinons \cite{spinhydro3}. Hence, akin to charge-doped graphene,  a ``spinon-doped" (realized by applying an in-plane magnetic field) Kagome spin liquid is expected to be another platform in which viscous effects in spin transport become evident.

 The rest of the paper is organized as follows: In Sec.~\ref{sec:Setup}, we describes the proposed protocol to detect the viscosity effects and the theoretical framework used in our analysis. Sec.~\ref{sec:Diffusive}  examines the spatial profiles of spin chemical potential and stray magnetic fields for the diffusive case in which the viscous effect is absent. Sec.~\ref{sec:Hydrodynamic} presents the results for the two hydrodynamic regimes and highlights how viscosity modifies the stray magnetic fields. We conclude with a summary and outlook in Sec.\ref{sec:Conc}.

\section{Setup and Method}
\label{sec:Setup}

\begin{figure}
    \centering
    \includegraphics[width=1.0\linewidth]{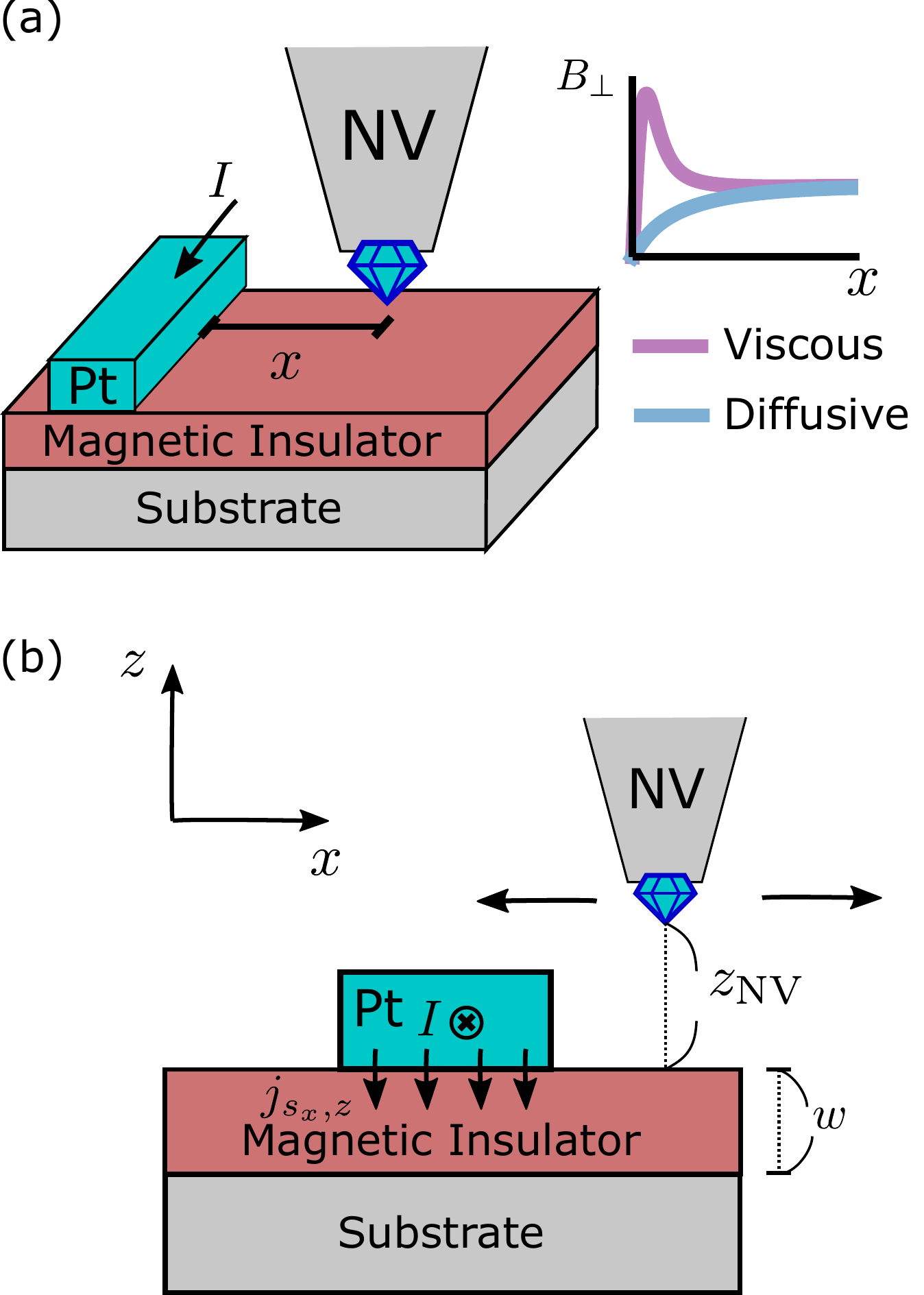}
    \caption{(a) Illustration of the proposed setup and our result. $B_{\perp}$ induced by DC spin transport features a sharp increase near the platinum strip used as a spin current injector in presence of viscosity. (b) The $xz$-plane cross section of the setup.}
    \label{fig:fig1}
\end{figure}

 The setup explored in this paper is illustrated in Fig.~\ref{fig:fig1}(a) and (b). A platinum strip is mounted on the smooth surface of a magnetic insulator film of width $w$ grown on a substrate. Owing to the strong spin Hall effect of platinum, flowing an electric current through the platinum strip along the $y$-direction generates a spin current polarized along the $x$-direction and flowing along the $z$-direction. This current, symbolized as $j_{s_{x},z}$ in Fig.~\ref{fig:fig1}(b), is injected into the magnetic insulator. If the magnetic insulator supports magnons, spinons, or any other coherent spin-carrying excitations, these injected spins can spread throughout the magnetic insulator bulk, leading to DC spin transport phenomena.

  Just as electron flow in a metal  is accompanied by an electric potential gradient, a DC spin current in a magnetic insulator induces a spatially varying spin chemical potential $\mu_{s_{x}}(\mathbf{x})$. This local spin chemical potential is, in turn, related to the spin magnetization density $n_{s_{x}}(\mathbf{x})$. At the level of linear response, one may assume the linear relation of the form $\delta n_{s_{x}}(\mathbf{x}) = C_{s} \mu_{s_{x}} $, where $C_{s} = dn_{s_{x}}/d \mu_{s_{x}}$ is the spin compressibility, and $\delta n_{s_{x}}(\mathbf{x}) = n_{s_{x}}(\mathbf{x}) - n_{s_{x},0}$ the change in $n_{s_{x}}(\mathbf{x})$ from its value $n_{s_{x},0}$ in the absence of an injected spin current. Here, $\mu_{s_{x}}$ is set to zero when no spin current is present.
  
  It is not clear if one can directly measure $n_{s_{x}}(\mathbf{x})$, or equivalently, $\mu_{s_{x}}(\mathbf{x})$ in the magnetic insulator bulk. However, the stray magnetic field profile that reflects the pattern of $n_{s_{x}}(\mathbf{x})$ and $\mu_{s_{x}}(\mathbf{x})$ can be directly probed using local magnetometers, which we assume to be a NV center throughout this paper. Hence, the goal of this paper is to study how viscous effects of hydrodynamic spin transport show up in the spatial distribution of the stray magnetic field. 
  
  The specific observable we focus on is the out-of-plane component of the magnetic field, $B_{\perp,z_{\text{NV}}}(x)$, where $z_{\text{NV}}$ is the vertical distance from the magnetic insulator thin film and $x$ the horizontal distance from the the spin current injector; see Fig.~\ref{fig:fig1}(a) and (b) for an illustration. The following subsections detail the formalism used to compute $\mu_{s_{x}}(\mathbf{x})$ and $B_{\perp,z_{\text{NV}}}(x)$ in each transport regime.

 \subsection{Spin current/potential profile in the diffusive regime}

  In the diffusive regime, spin transport is described by the following equations, where $\mathbf{j}_{s_{x}} = j_{s_{x},x} \hat{x} + j_{s_{x},y} \hat{y} + j_{s_{x},z} \hat{z}  $ is the spin current, $\sigma_{s,i}$ the spin conductivity along the direction $i$, and $\tau_{s}$ spin lifetime that arises due to spin non-conserving interactions:
\begin{equation}
\label{eq:diffusiveeq}
    j_{s_{x},i} = - \sigma_{s,i} \partial_{i}  \mu_{s_{x}}, \quad \frac{\partial  \delta  n_{s_{x}}}{ \partial t} + \nabla \cdot \mathbf{j}_{s_{x}} = -  \delta n_{s_{x}}/\tau_{s}.
\end{equation}
 Allowing $\sigma_{s,x}=\sigma_{s,y} \neq \sigma_{s,z}$, we take into account the anisotropy present in the case in which the magnetic insulators of interest are quasi-2D. Also, assuming that spontaneous or explicit symmetry breaking via an in-plane magnetic field reduces the approximate spin conservation symmetry to U$(1)$ along the $x$-axis, we ignore spin dynamics involving the $y$- and $z$-components.

 To further simplify the equations, one may assume that the setup is translationally symmetric along the $y$-direction and that the $n_{s}$ has no dependence on $y$. Similarly, we neglect time dependence of the equations in DC transport. Taking advantage of these two simplifications and the linear-response relation $\delta n_{s_{x}}(\mathbf{x}) = C_{s} \mu_{s_{x}}(\mathbf{x})$, we model the spatial profiles of $\mu_{s_{x}}$ and $\mathbf{j}_{s_{x}}$ with a solution to the 2D boundary value problem in the region $-\infty < x < \infty$ and $-w<z<0$ (recall that $w$ is the thickness of the magnetic insulator film), with the following partial differential equations:
 \begin{equation}
 \label{eq:diffusivepde}
 \begin{split}
     & \partial_{x}j_{s_{x},x}  +  \partial_{z}j_{s_{x},z} = - C_{s} \mu_{s_{x}} /\tau_{s} \\
     & j_{s_{x},x} = -\sigma_{s,x} \partial_{x}  \mu_{s_{x}}, \quad j_{s_{x},z} = -\sigma_{s,z} \partial_{z}  \mu_{s_{x}},
\end{split}
 \end{equation}
 satisfying the following boundary conditions:
 \begin{equation}
 \label{eq:diffusivebc}
      j_{s_{x},z}(x,z=-w) = 0, \quad j_{s_{x},z}(x,z=0) = -I_{s} \delta(x).
 \end{equation}
 The first boundary condition assumes that the substrate does not support spin transport and that spin current does not leak out from the magnetic insulator to the substrate. The second condition simplifies the platinum strip to be a delta function source of spin current located at $(x,z) = (0,0)$. 
 
   The standard separation of variables technique solves the above boundary value problem. For our purposes, it is convenient to apply a Fourier transform in the $x$-direction only. We define the following function:
  \begin{equation}
  \label{eq:Fourier}
      \tilde{\mu}_{s_{x},k}(z) = \int_{-\infty}^{\infty} dx \, e^{i k x}  \mu_{s_{x}}(x,z).
  \end{equation}
  As we will see later, the partial differential equation for $ \mu_{s_{x}}(x,z)$ in Eq.~\eqref{eq:diffusivepde} becomes the second-order ordinary differential equation for $\tilde{\mu}_{s_{x},k}(z)$ and can be solved straightforwardly.
 
 \subsection{Spin current/potential profile in the 3D hydrodynamic regime}
  
 We proceed similarly for the hydrodynamic regime: First, we present hydrodynamic equations that govern spin transport, valid at the linear response level. Then, we formulate the boundary value problem whose solution gives the spatial profiles of $\mu_{s_{x}}$ and $\mathbf{j}_{s_{x}}$. 

  In the regime of interest in this subsection, in addition to $S_{x}$ (as before, we assume that spontaneous magnetic ordering or an external in-plane magnetic field always kills the full $\text{SU}(2)$ spin rotation symmetry), momentum $\mathbf{P}$ emerges as an approximately conserved quantity. Accordingly, spin transport in the hydrodynamic regime is governed by the following approximate conservation laws where $\tau_{p}$ is the momentum lifetime, $\Pi_{ij}$ the energy-momentum tensor, and $i,j=x,y,z$:
  \begin{equation}
  \label{eq:hydrodynamiceq1}
      \frac{\partial  \delta  n_{s_{x}}}{ \partial t} + \nabla \cdot \mathbf{j}_{s_{x}} = -  \delta  n_{s_{x}}/\tau_{s}, \quad \frac{\partial P_{j}}{ \partial t} + \partial_{i} \Pi_{ij} = -P_{j}/\tau_{p}.
  \end{equation}
  We used the Einstein summation notation for the momentum conservation equations. As in Eq.~\eqref{eq:diffusiveeq}, the finite $\tau_{s}$ and $\tau_{p}$ account for spin-non conserving interactions and momentum-relaxing collisions, in a manner analogous to the relaxation time approximation in the Boltzmann equation framework. We neglect energy conservation equations which are typically included in hydrodynamic treatments, on the grounds that in DC spin transport experiments, the magnetic insulator thin film constantly exchanges heat with its environment so that the temperature $T$ throughout the film effectively remains constant.

 In addition to the conservation equation, the constitutive equations that relate $\mathbf{j}_{s_{x}}$ and $\Pi_{ij}$ to the hydrodynamic variables -- $\mathbf{P}$ and $\delta n_{s_{x}} = C_{s} \mu_{s_{x}}$ -- and their spatial derivatives fully specify the dynamics. We take the constitutive relations to be:
  \begin{equation}
  \begin{split}
  \label{eq:hydrodynamiceq2}
      & \mathbf{j}_{s_{x}} = \mathcal{N}_{s} \mathbf{P} - \sigma_{s,d} \nabla \mu_{s_{x}}\\
      & \Pi_{ij} = \left[ - \frac{\partial p}{\partial \mu_{s_{x}} }  \mu_{s_{x}}+ (\rho - \frac{2}{3}\eta) (\nabla \cdot \mathbf{P})  \right] \delta_{ij} + \eta (\partial_{i} P_{j} + \partial_{j} P_{i}).
  \end{split}
  \end{equation}
 The first relation encodes that spin current has two contributions: A ``hydrodynamic term" proportional to $\mathbf{P}$ in which coherent momentum carries spin current and a ``diffusive term" proportional to $\nabla \mu_{s_{x}}$. The second relation captures the standard features of fluid dynamics and Navier-Stokes equations: $p$ is pressure, $\rho$ the bulk viscosity, and $\eta$ the shear viscosity.

  In hydrodynamic approaches, constitutive relations are fixed by symmetries and the laws of thermodynamics \textit{without} invoking microscopic details. In the context of our setup for DC spin transport, the constitutive relations introduced above can be justified by the following assumptions:
\begin{itemize}
    \item Since we are interested in the linear response regime, we  retain only terms linear in the hydrodynamic variables. Similarly, since we are interested in the long-wavelength dynamics, we keep track of terms with zero or one spatial derivative only. 
    \item As mentioned in the beginning, spontaneous magnetic ordering or an in-plane magnetic field always breaks time-reversal symmetry $\mathcal{T}$ in our setup. However, if the Hamiltonian of the magnetic insulator respects a symmetry $\mathcal{C}: S_{x} \rightarrow - S_{x}$, then the combined operation $\mathcal{CT}$ may remain a good symmetry even after the symmetry breaking. We comment on the necessity of broken $\mathcal{T}$ in the hydrodynamic regimes and motivate $\mathcal{CT}$ in Appendix~\ref{app:timereversal}.
    
    We assume that $\mathcal{CT}$ serves as an effective time-reversal symmetry on the setup of our interest. Then, terms that contain no spatial derivative are ``reactive" parts of the relations which transform in the same way as the current on the left hand side of the equations Eq.~\eqref{eq:hydrodynamiceq2} under $\mathcal{CT}$. On the other hand, terms with one spatial derivative are ``dissipative" parts which should transform oppositely to the current under $\mathcal{CT}$ \cite{Chaikin1995}. 
    \item The system is isotropic and possesses a \text{SO}(3) spatial rotation symmetry. The constitutive relations should be compatible with the spatial rotation as well. 
\end{itemize}

  What we have in Eq.~\eqref{eq:hydrodynamiceq1} and Eq.~\eqref{eq:hydrodynamiceq2} are essentially equivalent to linearized Navier-Stokes equations for hydrodynamic charge transport upon replacing the spin density $\delta n_{s_{x}}$, chemical potential $\mu_{s_{x}}$, and current $\mathbf{j}_{s_{x}}$ with their charge analogues, and setting $\tau_{s} \rightarrow \infty$. However, we invoked somewhat different set of assumptions from ones used in the usual electron hydrodynamics. In particular, the lack of $\mathcal{T}$ and space-time symmetries \cite{spinhydro1,spinhydro3} such as Lorentz or Galilean symmetries \footnote{If one assumes Lorentz or Galilean symmetries, it is possible to fix the \textit{non-linear} part of the reactive terms as well. In particular, this non-linear part is responsible for convective terms in the Navier-Stokes equation. However, given that space-time symmetries are not generally valid in the systems of our interest \cite{spinhydro1,spinhydro3}, we restrict ourselves to linear terms which can be fixed without space-time symmetries.} lead us to make alternative assumptions to construct the constitutive relations.

 Having hydrodynamic equations and their justifications in hand, we now formulate the boundary value problem that determines the spatial profiles of $n_{s_{x}}$ and $\mathbf{j}_{s_{x}}$ in the hydrodynamic regime from Eq.~\eqref{eq:hydrodynamiceq1} and \eqref{eq:hydrodynamiceq2}. As we did for the diffusive case, we assume that the solution has no $y$- and $t$-dependence and $j_{s_{x},y} = 0$. Also, we previously noted that the term $-\sigma_{s,d} \nabla \mu_{s_{x}}$ in the first line of Eq.~\eqref{eq:hydrodynamiceq2} encodes diffusive contributions. Since we are primarily interested in exploring hydrodynamic effects, we omit this term. Under these assumptions, Eq.~\eqref{eq:hydrodynamiceq1} and \eqref{eq:hydrodynamiceq2} reduce to:
  \begin{equation}
  \label{eq:hydrodynamicpde}
  \begin{split}
      & \partial_{x} j_{s_{x},x} + \partial_{z} j_{s_{x},z} = - C_{s}  \mu_{s_{x}} /\tau_{s} \\
      & j_{s_{x},i=x,z} = -  \sigma_{s,\text{eff}} \partial_{i}  \mu_{s_{x}} + l_{v}^{2} (\partial_{x}^{2} + \partial_{z}^{2})  j_{s_{x},i}. 
  \end{split}
  \end{equation}
 In the second line, we simplified the momentum conservation equation by replacing $P_{i}$ with $\mathcal{N}_{s}^{-1} j_{s_{x},i}$ and introducing the following coefficients:
\begin{equation}
    \sigma_{s,\text{eff}} = \mathcal{N}_{s}^{-1} \tau_{p} \frac{\partial p}{\partial \mu_{s_{x}}} + (\rho + \frac{1}{3} \eta)  \frac{C_{s} \tau_{p}}{\mathcal{N}_{s} \tau_{s}} , \quad l_{v} = \sqrt{\tau_{p} \eta }.
\end{equation}
 $\sigma_{s,\text{eff}}$ is the coefficient that relates $j_{s_{x},i}$ to $\partial_{i} \mu_{s_{x}}$ and may be thought as an effective spin conductance in the hydrodynamic regime. $l_{v}$ is so-called Gurzhi length and quantifies the strength of the shear viscosity. The only difference between the diffusive equations in Eq.~\eqref{eq:diffusivepde} and the hydrodynamic equations in Eq.~\eqref{eq:hydrodynamicpde}, upon identifying $\sigma_{s,\text{eff}} = \sigma_{s,x} = \sigma_{s,z}$, is the presence of the term $l_{v}^{2} (\partial_{x}^{2} + \partial_{z}^{2}) j_{s_{x},i}$ that accounts for the shear viscosity effects.
 
 We intend to solve the system of partial differential equations Eq.~\eqref{eq:hydrodynamicpde} in the region $-\infty < x < \infty$ and $-w<z<0$, along with boundary conditions. We impose the same boundary condition as Eq.~\eqref{eq:diffusivebc} for the diffusive case and the following additional conditions on $j_{s_{x},x}$ required for solving Navier-Stokes-type equations:
\begin{equation}
\label{eq:hydrodynamicbc}
\begin{split}
    & j_{s_{x},x}(x,z=0,-w) =0 \quad \text{ or } \\
    &  \partial_{x} j_{s_{x},z}(x,z=0,-w) + \partial_{z} j_{s_{x},x}(x,z=0,-w) = 0.
\end{split}
\end{equation}
 The first boundary condition is the no-slip boundary condition that assumes due to the strong interface friction, the spin current at $z=0,-w$ should vanish. Meanwhile, the second boundary condition is the no-stress boundary condition which asserts $\Pi_{zx}=0$ at the interface due to the \textit{absence} of the friction. Both in fluid dynamics and electron hydrodynamics, no-slip boundary condition is the standard choice. However, in spin transport experiments the surface roughness can be much smaller due to high-quality thin film growth \footnote{We thank Aaron Hui for pointing out this possibility }. Hence, we will consider both boundary conditions in our analysis. 

 The partial differential equations in Eq.~\eqref{eq:hydrodynamicpde} and the boundary conditions in Eq.~\eqref{eq:diffusivebc} and Eq.~\eqref{eq:hydrodynamicbc} define the 2D boundary value problem we aim to solve for the 3D hydrodynamic regime. While the equations are more complicated than in the diffusive regime, they can still be solved with the same method: separation of variables technique via Fourier transformation of the $x$-coordinate. 

 \subsection{Spin current/potential profile in the 2D hydrodynamic regime}

  Here, we consider the 2D hydrodynamic regime in which the in-plane momenta $P_{x}$ and $P_{y}$ (but not $P_{z}$) emerge as good hydrodynamic variables. As mentioned at the introduction, the materials we primarily have in mind for this seemingly unusual regime are 2D quantum spin liquid candidates grown as stacks of 2D layers. In these materials, hydrodynamic regimes may emerge from rapid thermalization of spinons via strong gauge-mediated interactions within a 2D layer \cite{spinhydro3}. If Umklapp or impurity scattering is suppressed, the in-plane momenta can emerge as approximately conserved quantities. 
  
  The most straightforward scenario in which in-plane momentum-relaxing scattering can be suppressed is when spinons have Dirac-like dispersions, as previous theoretical works on Kagome antiferromagnets suggest. On the other hand, the interlayer coupling is much weaker \cite{Zou2016}, and the spinon dispersion is essentially flat in the $k_{z}$ direction. As a result, spinons are highly susceptible to Umklapp scattering that leads to a rapid loss of $P_{z}$. Slow diffusion of spins and in-plane momenta mediated by weak interlayer coupling will govern out-of-plane dynamics instead of $P_{z}$. While 2D spin liquids provide the most dramatic example, we note that the 2D hydrodynamic regime may also appear in quasi-2D ordered magnets in which interlayer exchange interaction $J_{\text{ex}, \perp}$ is much weaker than intralayer interaction $J_{\text{ex}, \parallel}$.

  As the previous two cases, the hydrodynamic equations that govern spin transport are given as conservation equations and constitutive relations (the Greek alphabets $\alpha, \beta =x,y$ only span the in-plane indices, while $i=x,y,z$ spans all spatial indices as before):
\begin{equation}
  \label{eq:ahydrodynamiceq1}
      \frac{\partial  \delta  n_{s_{x}}}{ \partial t} + \nabla \cdot \mathbf{j}_{s_{x}} = -  \delta  n_{s_{x}}/\tau_{s}, \quad \frac{\partial P_{\alpha}}{ \partial t} + \partial_{i} \Pi_{i\alpha} = -P_{\alpha}/\tau_{p},
  \end{equation}
  \begin{equation}
  \begin{split}
  \label{eq:ahydrodynamiceq2}
      & j_{s_{x},\alpha} = \mathcal{N}_{s} P_{\alpha} - \sigma_{s,d} \partial_{\alpha} \mu_{s_{x}} \\ 
      & \Pi_{\alpha\beta} = \left[ - \frac{\partial p}{\partial \mu_{s_{x}}} \mu_{s_{x}} + (\rho - \eta) \partial_{\alpha} P_{\alpha}  \right] \delta_{ij} + \eta (\partial_{\alpha} P_{\beta} + \partial_{\beta} P_{\alpha}) \\ 
      & j_{s_{x},z} = - \sigma_{s,\perp} \partial_{z} \mu_{s_{x}}, \quad  \Pi_{z\alpha} = -\eta_{\perp} \partial_{z} P_{\alpha}.
  \end{split}
  \end{equation}

  The first two constitutive relations in Eq.~\eqref{eq:ahydrodynamiceq2} are simply the 2D analogs of Eq.~\eqref{eq:hydrodynamiceq2}, and the physical interpretations of the coefficients $\mathcal{N}_{s}$, $\sigma_{s,d}$ $p$, $\rho$,$\eta$ remain unchanged. Meanwhile, the relations in the third line encode diffusion of in-plane momenta and spins along the $z$ direction; $\eta_{\perp}$ and $\sigma_{s,\perp}$ can be understood as the corresponding diffusion constants. These constitutive relations are fixed with the same conditions we imposed for the 3D hydrodynamic constitutive relation except for the spatial isotropy; the $\text{SO}(3)$ spatial rotation symmetry assumed for the 3D hydrodynamic regime is replaced here by $\text{SO}(2)$ rotation and mirror symmetry $M_{x}: x \rightarrow -x$, or equivalently, $M_{y}: y \rightarrow -y$ \footnote{If $M_{x}$ and space-time symmetries are absent, a term linear in $\hat{z} \times \mathbf{P}$ is allowed in the constitutive relation for $\mathbf{j}_{s_{x}}$. While we impose mirror symmetry to forbid this term, it would be interesting to investigate whether this type of term can emerge in a more microscopic treatment when $M_{x}$ is broken.}.

 Using Eq.~\eqref{eq:ahydrodynamiceq1}, Eq.~\eqref{eq:ahydrodynamiceq2}, and the same assumptions we made to formulate the boundary value problem in the 3D hydrodynamic regime, one can show that solving the following partial differential equation gives the spatial map of $\mathbf{j}_{s_{x}}$ and $\mu_{s_{x}}$ in the 2D hydrodynamic regime ($l_{v} = \sqrt{(\rho + \eta) \tau_{p}}$, $l_{v,\perp} = \sqrt{\eta_{\perp} \tau_{p}}$, and $\sigma_{s,\parallel} = \mathcal{N}_{s}^{-1} \tau_{p} \frac{\partial p}{\partial n_{s_{x}}}$):
\begin{equation}
\begin{split}
& \partial_{x} j_{s_{x},x} - \sigma_{s,\perp} \partial_{z}^{2}  \mu_{s_{x}}  = - \mu_{s_{x}} C_{s} / \tau_{s} \\
& j_{s_{x},x} = -\sigma_{s,\text{eff}} \partial_{x} \mu_{s_{x}}  + l_{v}^{2} \partial_{x}^2 j_{s_{x},x} + l_{v,\perp}^{2} \partial_{z}^2 j_{s_{x},x}.
\end{split}
\end{equation}
The relevant boundary conditions are: 
\begin{equation}
\label{eq:ahydrodynamicbc}
\begin{split}
& j_{s_{x},z}(x,z=-w) = - \sigma_{s,\perp} \partial_{z} \mu_{s_{x}}(x,z=-w) = 0, \\
& j_{s_{x},z}(x,z=0) = - \sigma_{s,\perp} \partial_{z}\mu_{s_{x}}(x,z=0) = -I_{s}  \delta(x), \\ 
& j_{s_{x},x}(x,z=0,-w) =0 \text{\quad or \quad }  \\ 
& \Pi_{zx}(x,z=0,-w)  \propto \partial_{z} j_{s_{x},x}(x,z=0,-w) = 0.
\end{split}
\end{equation}
 
\subsection{Stray magnetic field from DC spin transport}

 We have shown how to obtain the spin chemical potential profile $\mu_{s_{x}}(x,z)$, induced by spin current injection, as a solution to the boundary value problem defined over the region $-\infty<x<\infty$ and $-w<z<0$. We now elucidate how to compute the magnetic field sensed by an NV center located above the magnetic insulator film at the coordinate $(x, z_{\text{NV}})$.

 Assuming that magnetic moments in magnetic insulators interact with spins in the NV center through dipolar interactions, the stray magnetic field can be simply computed by integrating contributions in the region $-\infty< x,y <\infty$ and $-w<z<0$ ($\mu_{0}$ is the vacuum permeability):
\begin{equation}
\begin{split}
    \mathbf{B}(x, z_{\text{NV}}) = \frac{\mu_{0}}{4\pi} \int_{-\infty}^{\infty} dx' \int_{-\infty}^{\infty} dy' \int_{-w}^{0} dz'\,  n_{s_{x}}(x',z')   \\
    \frac{3 (x-x') \left[(x-x') \hat{\mathbf{x}} + (y-y')\hat{\mathbf{y}} +(z_{\text{NV}}-z')\hat{\mathbf{z}} \right]}{\left[ (x-x')^{2} + (y-y')^{2} +(z_{\text{NV}}-z')^{2} \right]^{5/2} } \\
    - \left[ (x-x')^{2} + (y-y')^{2} +(z_{\text{NV}}-z')^{2} \right]^{-3/2} \hat{\mathbf{x}}.
\end{split}
\end{equation}
 As mentioned at the beginning of this section, we will focus on $B_{\perp, z_{\text{NV}}}(x)$ which admits a particularly simple physical interpretation. By explicitly integrating over $y'$ and applying integration by parts formula on $x'$, one can obtain:
\begin{equation}
\begin{split}
    & B_{\perp, z_{\text{NV}}}(x) \\
    & = \frac{\mu_{0}}{\pi} \int_{-\infty}^{\infty} dx'  \int_{-w}^{0} dz'\, n_{s_{x}}(x',z')    \frac{  (x-x')(z_{\text{NV}}-z')}{\left[ (x-x')^{2}  +(z_{\text{NV}}-z')^{2} \right]^{2} }     \\
    & = \frac{\mu_{0} C_{s} }{\pi} \int_{-\infty}^{\infty} dx'  \int_{-w}^{0} dz'\,   \\
    & \quad \quad \quad \quad  \partial_{x'} \mu_{s_{x}}(x',z')\frac{   (z_{\text{NV}}-z')}{ (x-x')^{2}  +(z_{\text{NV}}-z')^{2}}.   
\end{split}
\end{equation}
 In the final line, we used the relation $\partial_{x'} n_{s_{x}}(x',z') = \partial_{x'} \delta n_{s_{x}}(x',z') = C_{s} \partial_{x'} \mu_{s_{x}}(x',z')$. One can see that $B_{\perp, z_{\text{NV}}}(x)$ is a weighted average of $\partial_{x} \mu_{s_{x}}$; the weight is a Lorentzian function centered at $x$ with a width set by $z_{\text{NV}}-z'$. This is the main result of this subsection.

 From a practical standpoint, it is often more useful to work with the Fourier-transformed potential $\tilde{\mu}_{s_{x},k}(z') = \int_{-\infty}^{\infty} dx' \, e^{i k x'}  \delta \mu_{s_{x}}(x',z')$ for which closed-form solutions in terms of elementary functions exist. Using the integral identity $\int^{\infty}_{-\infty} dx' \, e^{-i k x' } \frac{(z_{\text{NV}}-z')}{(x-x')^{2} + (z_{\text{NV}}-z')^{2}}  = \pi e^{-i k x } \text{sgn}(z_{\text{NV}}-z') e^{-|k||z_{\text{NV}}-z'|}$, $B_{\perp, z_{\text{NV}}}(x)$ can be alternatively written as:
\begin{equation}
\label{eq:Bperpk}
\begin{split}
    & B_{\perp, z_{\text{NV}}}(x) \\
    & = \frac{\mu_{0} C_{s}}{2\pi} \int_{-\infty}^{\infty} dk \int_{-w}^{0} dz'\, (-ik) e^{-i k x' }   \tilde{\mu}_{s_{x},k}(z') e^{-|k|(z_{\text{NV}}-z')}\\
    & = -\frac{\mu_{0} C_{s} }{\pi} \int_{0}^{\infty} dk \int_{-w}^{0} dz'\, k  \sin k x \,    \tilde{\mu}_{s_{x},k}(z') e^{-k(z_{\text{NV}}-z')}.
\end{split}
\end{equation}
In the final step, we used the fact that our setup is symmetric under $x \rightarrow -x$, so $\tilde{\mu}_{s_{x},k}(z') =  \tilde{\mu}_{s_{x},-k}(z')$. While the $z'$ integral can be only evaluated after plugging in the explicit form of $\tilde{\mu}_{s_{x},k}(z')$, we will see that it is fairly simple to carry out for $\tilde{\mu}_{s_{x},k}(z')$ computed in this paper. Hence, one can express $B_{\perp, z_{\text{NV}}}(x)$ in terms of a single Fourier integral which can be efficiently evaluated numerically.

\section{Diffusive regime}
\label{sec:Diffusive}

 Here, we study $B_{\perp, z_{\text{NV}}}(x)$ in the diffusive regime. Given that the $x$ coordinate of the local magnetometers is highly tunable, we will focus on the one-dimensional profile of $B_{\perp, z_{\text{NV}}}(x)$ with fixed $z_{\text{NV}}$. Although tuning $z_{\text{NV}}$ is more difficult, it is still experimentally feasible to measure $B_{\perp, z_{\text{NV}}}(x)$ for a few different values of $z_{\text{NV}}$. Hence, we also explore how the one-dimensional profile of $B_{\perp, z_{\text{NV}}}(x)$ changes as $z_{\text{NV}}$ is varied.

 \subsection{Solution to the boundary value problem}
\label{sec:solutiondiffusive}
 First, we briefly state how to obtain the spin potential profile inside the magnetic insulator $ \mu_{s_{x}}(x,z)$ by solving the boundary value problem. Using the Fourier-transformed function $\tilde{\mu}_{s_{x},k}(z)$ defined in Eq.~\eqref{eq:Fourier}, one can express the partial differential equation as the following 2D ordinary differential equation, where $\xi_{s} = \sqrt{\tau_{s} \sigma_{s,x}/C_{s}} $ is the spin diffusion length along the $x$-direction and $0< r_{\sigma} = \sigma_{s,z}/ \sigma_{s,x} <1$ is the ratio characterizing the anisotropy of the diffusion constants:
 \begin{equation}
     \partial_{z}^{2} \tilde{\mu}_{s_{x},k} - r_{\sigma}^{-1} (k^2 + \xi_{s}^{-2}) \tilde{\mu}_{s_{x},k} =0. 
 \end{equation}
 Upon defining $q_{s} =  \sqrt{r_{\sigma}^{-1} (k^2 + \xi_{s}^{-2})}$, the general solution take the following form:
\begin{equation}
\label{eq:nsdiffusive}
    \tilde{\mu}_{s_{x},k}(z) = A(k) \cosh q_{s}(z+w) + B(k) \sinh q_{s}(z+w).
\end{equation}
$A(k)$ and $B(k)$ can be determined by matching the boundary conditions given in Eq.~\eqref{eq:diffusivebc}. Upon doing so, one obtains:
\begin{equation}
 \tilde{\mu}_{s_{x},k}(z)= \frac{I_{s}}{\sigma_{s,z}} \frac{\cosh q_{s}  (z+w) }{q_{s}  \sinh q_{s}  w}.
\end{equation}
Plugging in the above solution to Eq.~\eqref{eq:Bperpk} and performing the $z'$ integral, one obtains the following formula:

\begin{equation}
\label{eq:DiffusiveBz}
\begin{split}
   B_{\perp, z_{\text{NV}} }(x) & =   -\frac{\mu_{0} C_{s} I_{s}}{\pi  \sigma_{s,z}}  \int_{0}^{\infty}  dk \, \frac{k \sin k x e^{-k z_{\text{NV}}}}{q_{s} \left[1 - e^{ - 2 q_{s} w } \right]} \\
   & \left[ \frac{ 1- e^{-(k+q_{s})w }}{k + q_{s} } + \frac{e^{ - 2 q_{s} w } - e^{-(k+q_{s})w} }{k - q_{s} } \right]
\end{split}
 \end{equation}

\subsection{Asymptotics of $B_{\perp, z_{\text{NV}} }(x)$}
\label{sec:aympdiffusive}

It is not clear to us whether it is possible to integrate over $k$ analytically in Eq.~\eqref{eq:DiffusiveBz} to obtain an insightful closed-form expression for $B_{\perp, z_{\text{NV}} }(x)$ in terms of elementary functions. Its Fourier representation is nevertheless powerful enough to extract the asymptotic behaviors at small $x$ and large $x$. 

 To see how this works, first observe that one can always write $B_{\perp, z_{\text{NV}} }(x)$ in Eq.~\eqref{eq:DiffusiveBz} into a form $B_{\perp, z_{\text{NV}} }(x) = \int_{0}^{\infty}d k \, \left(e^{-k (z_{\text{NV}} + i x)} - e^{-k (z_{\text{NV}} - i x)} \right)  I(k)$.  The key input is that $I(k)$, originally defined only for real $k$, can be analytically continued to the entire half-complex plane $\text{Re}(k)>0$ in a way that $I(k)$ is analytic throughout this region. Hence, one can deform the integration paths in a way that $k (z_{\text{NV}} + i x)$ (or $k (z_{\text{NV}} - i x)$) is always real. 

  The integrand along the deformed lines decays exponentially as $e^{-k' \sqrt{x^{2} + z_{\text{NV}}^{2}}}$, with $k'$ parameterizing the deformed lines. Then, one can readily see that the large $x$ behavior is governed by the small-$k$ behavior of $I(k) \sim ik$. This asymptotic behavior of $I(k)$ is valid when $x$ is much smaller than two length scales that appear in the expression for $I(k)$: $r_{\sigma}^{-1/2} w$ and $\xi_{s}$. By approximating $I(k)$ to be $ik$ and integrating , one obtains:
\begin{equation}
    B_{\perp, z_{\text{NV}} }(x) \sim \frac{1}{x^{2}} \quad \quad (x \gg r_{\sigma}^{-1/2} w, \xi_{s} ).
\end{equation}
One interesting thing to note is that, the $x^{-2}$ tail at large $x$ \textit{has no dependence} on $z_{\text{NV}}$.

 Similarly, the small-$x$ behavior can be inferred from the large-$k$ behavior of the integrand. $I(k) \sim ik^{-1}$ at large $k$; when $x$ is small, the integral along the deformed lines is dominated by $1/k$ tail at large $k$. Hence, by approximating $I(k)$ to be $ik^{-1}$ and doing the integral, one can obtain the following small $x$ asymptotics:
\begin{equation}
    B_{\perp, z_{\text{NV}} }(x)  \sim \tan^{-1} \frac{x}{z_{\text{NV}}} \quad (x \ll r_{\sigma}^{-1/2} w, \xi_{s} ).
\end{equation}
Note that if $z_{\text{NV}} \ll r_{\sigma}^{-1/2} w, \xi_{s}$, this asymptotics is valid even when $x$ is comparable to or larger than $z_{\text{NV}}$. This behavior shows that for small $x$, $B_{\perp, z_{\text{NV}} }(x)$ increases in a $z_{\text{NV}}$-dependent fashion and approaches a value \textit{independent} of $z_{\text{NV}}$.

 Assuming that $B_{\perp, z_{\text{NV}} }(x)$ smoothly interpolates between the two asymptotics, we infer the shape of $B_{\perp, z_{\text{NV}} }(x)$ as the following: As $x$ moves away from $0$,  $B_{\perp, z_{\text{NV}} }(x)$ initially increases, saturates to a fixed value, and eventually decays as $x^{-2}$ at the largest length scale. Only the initial increase in $B_{\perp}$ is $z_{\text{NV}}$-dependent.

\subsection{Numerical result}

\begin{figure}
    \centering
    \includegraphics[width=1.0\linewidth]{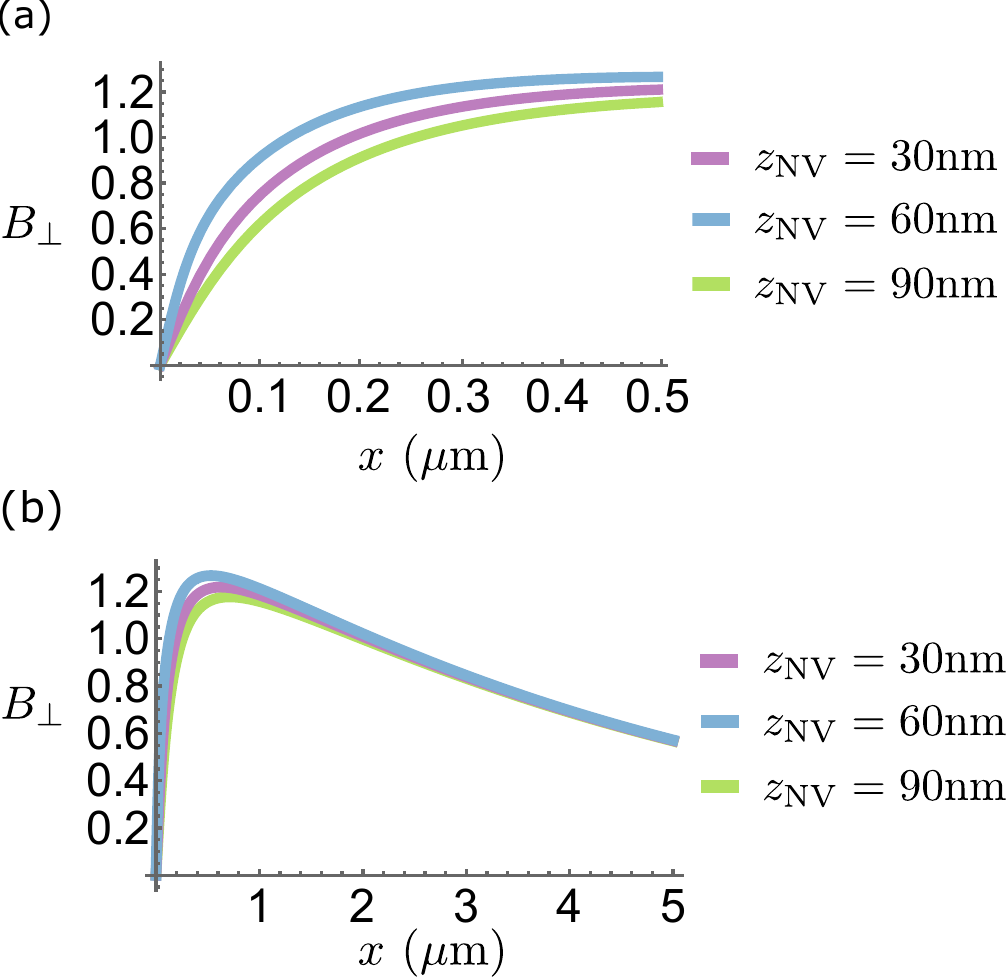}
    \caption{$B_{\perp, z_{\text{NV}} }(x)$ as a function of distance $x$ away from the injector in the diffusive regime. (a) $B_{\perp}$ close to the injector (b) $B_{\perp}$ far away from the injector.  Parameter values utilized to generate the two plots are: $r_{\sigma} = 1$, $\xi_{s} = 5 \text{$\mu$m}$, and $w=100\text{nm}$.}
    \label{fig:fig2}
\end{figure}

 While the asymptotic analysis of $B_{\perp}(x, z_{\text{NV}})$ allows us to fairly convincingly infer its shape, for a more sure-fire check, we turn to numerics. We numerically evaluate the integral in Eq.~\eqref{eq:DiffusiveBz} and plot its $x$-dependence.

  Fig.~\ref{fig:fig2} shows $B_{\perp, z_{\text{NV}} }(x)$ calculated for different values of $z_{\text{NV}}$'s to show how the profile depends on the NV center-sample distance. Parameter choice used here is $r_{\sigma} = 1$, $\xi_{s} = 5 \text{$\mu$m}$, and $w=100\text{nm}$. The values of $w$ and $\xi_{s}$ we used here are in line with values from the spin transport experiments on YIG and $\alpha$-Fe$_2$O$_3$ \cite{spint1,spint2}.

  Fig.~\ref{fig:fig2}(a) show the behavior close to the injector, while Fig.~\ref{fig:fig2}(b) shows the long-distance behavior. These plots pretty much numerically confirm the behavior we proposed from the asymptotic analysis. As one can see in Fig.~\ref{fig:fig2}(a), as one goes away from the injector, $B_{\perp}$ initially increases; the slope of this increase is $z_{\text{NV}}$-dependent. However, it saturates to a value which has a weak $z_{\text{NV}}$-dependence. As shown in Fig.~\ref{fig:fig2}(b), at the long distance where $x \sim \xi_{s}$, $B_{\perp}$ slowly decays toward zero. We checked that the same trend can be observed for different choices of $w$ and $r_{\sigma}\leq 1$ as well.

  \begin{figure}
    \centering
    \includegraphics[width=1.0\linewidth]{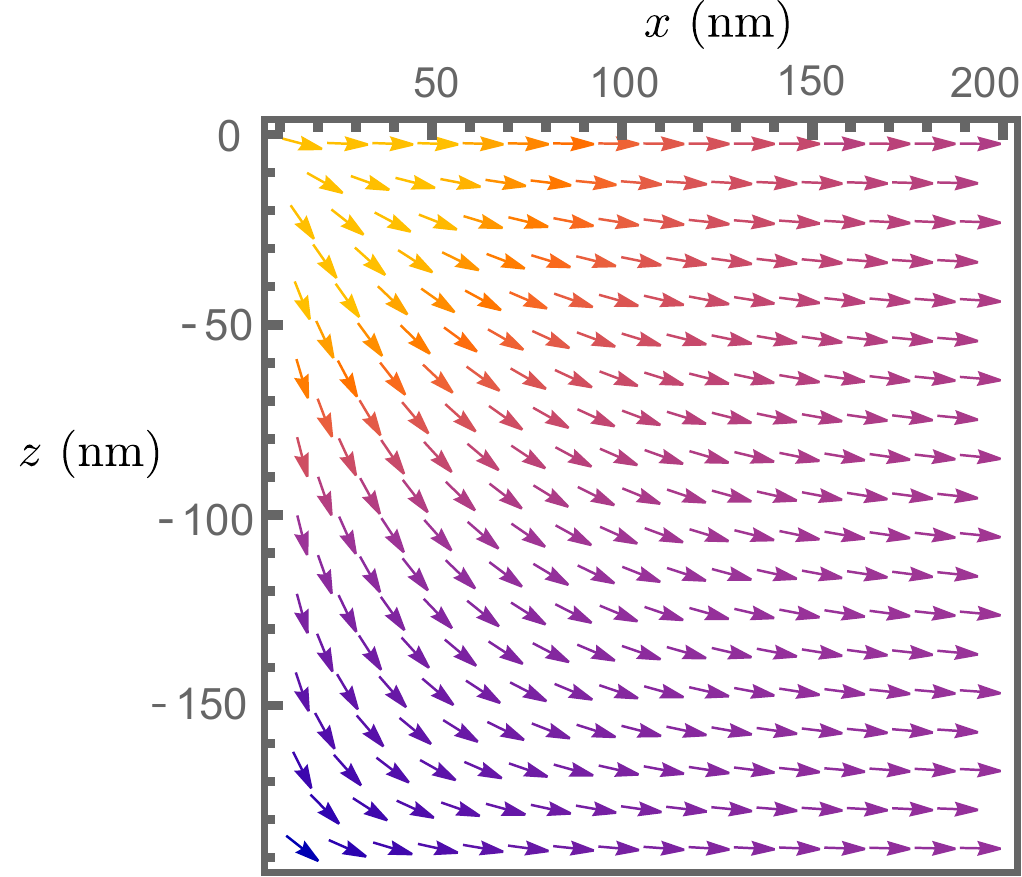}
    \caption{Current profile near the spin-current injector at $(x,z)=(0,0)$. $r_{\sigma} = 1$, $\xi_{s} = 5 \text{$\mu$m}$, and $w=200\text{nm}$ are used to generate the plot.}
    \label{fig:fig3}
\end{figure}

  It is instructive to build a more heuristic understanding of these behaviors. For this purpose, we plot the current profile obtained from our solution in Fig.~\ref{fig:fig3}. One can see that near the injector at $(x,z) = (0,0) $, the current is mostly vertical. However, as one moves away from the injector, the spin current makes a 90$\degree$ turn. In the diffusive regime, $\partial_{x} \mu_{s_{x}}$ is directly proportional to $j_{s_{x},x}$. Therefore, the initial increase in $B_{\perp}$ reflects an increase in $j_{s_{x},x}$ due to the turn the spin current is making.

  Meanwhile, as one moves farther away from the injector, the flow of the spin current becomes effectively one-dimensional. At $x \ll \xi_{s}$, the spin current is mostly conserved, and $j_{s_{x},x}$ is almost a constant. However, at $x \sim \xi_{s}$, the current decays exponentially due to spin-non-conserving processes. These two behaviors are also reflected in $B_{\perp, z_{\text{NV}} }(x)$  where $B_{\perp}$ saturates and then decays. However, due to the power-law nature of the dipolar interaction, the long-distance decay is $x^{-2}$ instead of exponential as the decay of the current profile would be.

\section{Hydrodynamic regime}
\label{sec:Hydrodynamic}

 Having established the spatial profile of $B_{\perp}$ in the diffusive regime, we now take a look into $B_{\perp}$ in the hydrodynamic regime, with a particular focus on how viscosity modifies the profile. First, we investigate the 3D hydrodynamics regime with no-stress boundary condition at the interface, where the solution to the boundary value is simple enough to perform the similar asymptotic analysis we did for the diffusive case. Then, we numerically explore other cases (no-slip boundary condition, 2D hydrodynamics) and demonstrate that, provided viscosity is sufficiently strong, the key signatures of the viscosity remain evident.

\subsection{Solution in the 3D hydrodynamic regime and its asymptotic behavior}

 Here, as in the diffusive case, we discuss how to solve the boundary value problem posed by Eq.~\eqref{eq:hydrodynamicpde} and \eqref{eq:hydrodynamicbc} and use its solution for the free-stress boundary condition to study the small-$x$ and large-$x$ asymptotic behavior.

 Eq.~\eqref{eq:hydrodynamicpde} is a system of partial differential equations for three different functions: $(j_{s_{x},x}, j_{s_{x},z}, \mu_{s_{x}})$. However, using an ansatz similar to the stream function commonly used in the 2D incompressible fluid dynamics, one can easily decouple the system of differential equations. 

 Instead of working directly the current vector, we introduce two scalar functions $\phi(x,z)$ and $\psi(x,z)$ to express the current as:
 \begin{equation}
 \begin{split}
     j_{s_{x},x} = \partial_{x} \phi(x,z) + \partial_{z} \psi(x,z) \\
     j_{s_{x},z} = \partial_{z} \phi(x,z) - \partial_{x} \psi(x,z).
 \end{split}
 \end{equation}
 In the usual stream function method, $\phi(x,z)$ should be zero due to the incompressibility condition -- the divergence of the fluid velocity vector vanishes. However, in our case, due to spin non-conserving processes, one cannot assume that the fluid is incompressible. Thus, we retain $\phi(x,z)$ which may be thought as taking account of the compressibility.

  Using this ansatz and taking curl and divergence of the second partial differential equations in Eq.~\eqref{eq:hydrodynamicpde}, one can show that the above ansatz reorganizes the partial differential equations into the following, where $\xi_{s,\text{eff}} = \sqrt{\sigma_{s,\text{eff}} \tau_{s}C_{s}^{-1} + l_{v}^{2}}$ plays the role of the spin diffusion length in this regime:
  \begin{equation}
  \begin{split}
  & \nabla^{2} (\nabla^{2} \psi - l_{v}^{-2}  \psi) = 0 \\
  & \nabla^{2} (\nabla^{2} \phi - \xi_{s,\text{eff}}^{-2}  \phi) = 0 \\
  & \nabla^{2} (\phi +\frac{\sigma_{s,\text{eff}}}{1 - l_{v}^{2} \xi_{s,\text{eff}}^{-2} }\mu_{s_{x}}) = 0.
  \end{split}
  \end{equation}
 Neglecting the overall $\nabla^{2}$ in each equations and introducing the Fourier transformation $\tilde{\psi}_{k}(z) = \int_{-\infty}^{\infty} dx \, e^{ikx}  \psi(x,z)$ and $\tilde{\phi}_{k}(z) = \int_{-\infty}^{\infty} dx \, e^{ikx}  \phi(x,z)$, the ansatz takes the following form:
 \begin{equation}
 \label{eq:3Dgeneral}
 \begin{split}
     & \tilde{\psi}_{k}(z) = A(k) \cosh \sqrt{k^{2} + \tilde{\xi}_{s}^{-2}} z +  B(k) \sinh \sqrt{k^{2} + \tilde{\xi}_{s}^{-2}} z \\
    & \tilde{\phi}_{k}(z) = C(k) \cosh \sqrt{k^{2} +  l_{v}^{-2}} z  + D(k) \sinh \sqrt{k^{2} +  l_{v}^{-2}} z  \\
    & \tilde{\mu}_{s_{x},k}(z) = \frac{1 - l_{v}^{2} \xi_{s,\text{eff}}^{-2} }{ \sigma_{s,\text{eff}} } \tilde{\phi}_{k}(z).
 \end{split}    
 \end{equation}
 One can show that matching the boundary condition in Eq.~\eqref{eq:hydrodynamicbc} uniquely specifies $A(k)$, $B(k)$, $C(k)$ and $D(k)$, fully solving the boundary value problem. Below, we present the solution for the no-stress case and its asymptotic behavior; the solution for the no-slip boundary condition is given in Appendix~\ref{app:noslip}. 

 For the free stress boundary condition case, the solution is simple enough, and $\tilde{\mu}_{s_{x},k}(z)$ is ($q_{s} = \sqrt{k^{2} + \xi_{s,\text{eff}}^{-2}}$):
 \begin{equation}
     \tilde{\mu}_{s_{x},k}(z) = \frac{I_{s} }{\sigma_{s,\text{eff}}}\left( 1 - l_{v}^{2} \xi_{s,\text{eff}}^{-2} \right)  \frac{(2k^{2}l_{v}^{2} + 1 )\cosh q_{s} (z+w) }{  q_{s} \sinh q_{s} w}.
 \end{equation}
 As before, plugging this solution to Eq.~\eqref{eq:Bperpk} gives the function of our interest:
 \begin{equation}
 \label{eq:Bperphydro}
     \begin{split}
   &B_{\perp, z_{\text{NV}} }(x)  =  - \frac{\mu_{0} C_{s} I_{s} }{2\pi \sigma_{s,\text{eff}}}\left( 1 - l_{v}^{2} \xi_{s,\text{eff}}^{-2} \right) \\
   & \quad \quad\int_{0}^{\infty} dk  \, \frac{(2k^{2}l_{v}^{2} + 1 ) k \sin k x e^{-k z_{\text{NV}}}}{q_{s} \left[1 - e^{ - 2 q_{s} w } \right]} \\ 
   & \quad \quad \quad\quad\left[ \frac{ 1- e^{-(k+q_{s})w }}{k + q_{s} } + \frac{e^{ - 2 q_{s} w } - e^{-(k+q_{s})w} }{k - q_{s} } \right].
\end{split}
 \end{equation}
 One can readily see that if $l_{v} =0$, $\tilde{n}_{s_{x},k}(z)$ and $B_{\perp}(x, z_{\text{NV}})$ reduce to the diffusive regime result in Eq.~\eqref{eq:nsdiffusive} and Eq.~\eqref{eq:DiffusiveBz}, up to an overall constant. In fact, the only difference of Eq.~\eqref{eq:Bperphydro} from the diffusive regime solution Eq.~\eqref{eq:DiffusiveBz} is (modulo an overall constant) the extra factor $(2k^{2}l_{v}^{2} + 1 ) $ in the integrand.

 Hence, one can extract the asymptotic behaviors using the same technique employed in Sec.~\ref{sec:aympdiffusive}. We find: 
 \begin{equation}
 \begin{split}
 & B_{\perp, z_{\text{NV}} }(x)  \sim \frac{1}{x^{2}}  \quad (x \gg  w, \xi_{s}  ) \\
 & B_{\perp, z_{\text{NV}} }(x)  \sim \tan^{-1} \frac{x}{z_{\text{NV}}}  +  \frac{l_{v}^{2}}{z_{\text{NV}}^{2}} \frac{x z_{\text{NV}}^{-1} }{(x^{2} z_{\text{NV}}^{-2} + 1 )^{2}} \quad (x \ll  w, \xi_{s} ) .
 \end{split}
 \end{equation}

 One can see that viscosity does not affect the long-distance behavior. However, near the injector, the viscosity introduce an extra term in the asymptotic expansion which sharply increases from $x=0$ to its maximum at $x \approx 0.58 z_{\text{NV}}$ and then decays as $\sim x^{-3} $. The strength of this contribution is proportional to $l_{v}^{2}$ and $z_{\text{NV}}^{-2}$, making it more pronounced at higher viscosity and closer NV-sample distances. In particular, the sharp $z_{\text{NV}}$-dependence distinguishes the viscosity-induced peak from features that also appear in the diffusive regime.
 
\subsection{Numerical result for the 3D hydrodynamic regime}

 \begin{figure}
    \centering
    \includegraphics[width=1.0\linewidth]{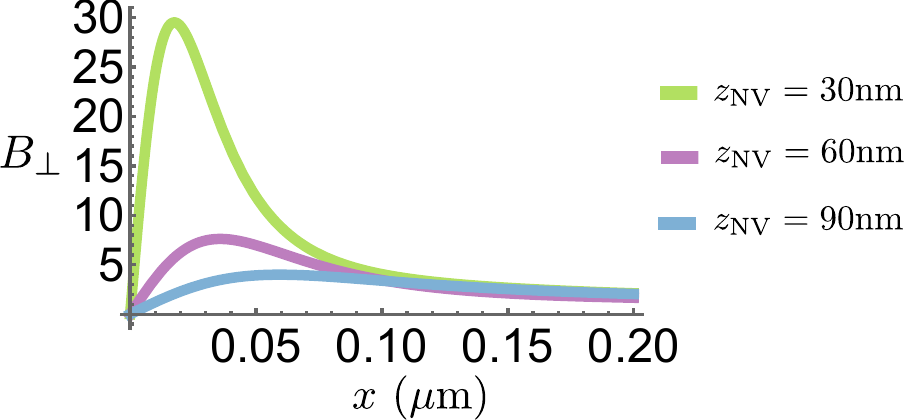}
    \caption{$B_{\perp}$ as a function of distance from the injector $x$; $\xi_{s,\text{eff}} = 50 \text{$\mu$m}$, $w=100\text{nm}$, and finally $l_{v}=200\text{nm}$. }
    \label{fig:fig4}
\end{figure}

Our asymptotic analysis already concretely shows, at least for the 3D isotropic hydrodynamic regime with the no-stress boundary condition, that a sharp peak in $B_{\perp}$ highly sensitive to $z_{\text{NV}}$ is the key signature of the viscous spin transport. To further verify this behavior and examine its robustness under different boundary conditions, we now turn to numerical simulation.

 We show the numerical result for no-stress boundary condition in Fig.~\ref{fig:fig4} with parameters $\xi_{s,\text{eff}} = 5 \text{ $\mu$m}$, $w=100\text{nm}$, and $l_{v}=200\text{nm}$. While we do not have a good estimate of $l_{v}$ in spin transport, we note that experiments in room-temperature graphene have estimated $l_{v} \sim 2 \text{$\mu$m}$\cite{visual}. As we predicted by the asymptotic analysis, one can indeed see that a pronounced peak appears near the injector and that its height exhibits strong dependence on $z_{\text{NV}}$, reinforcing the interpretation that this structure reflects a viscous response.

 \begin{figure}
    \centering
    \includegraphics[width=1.0\linewidth]{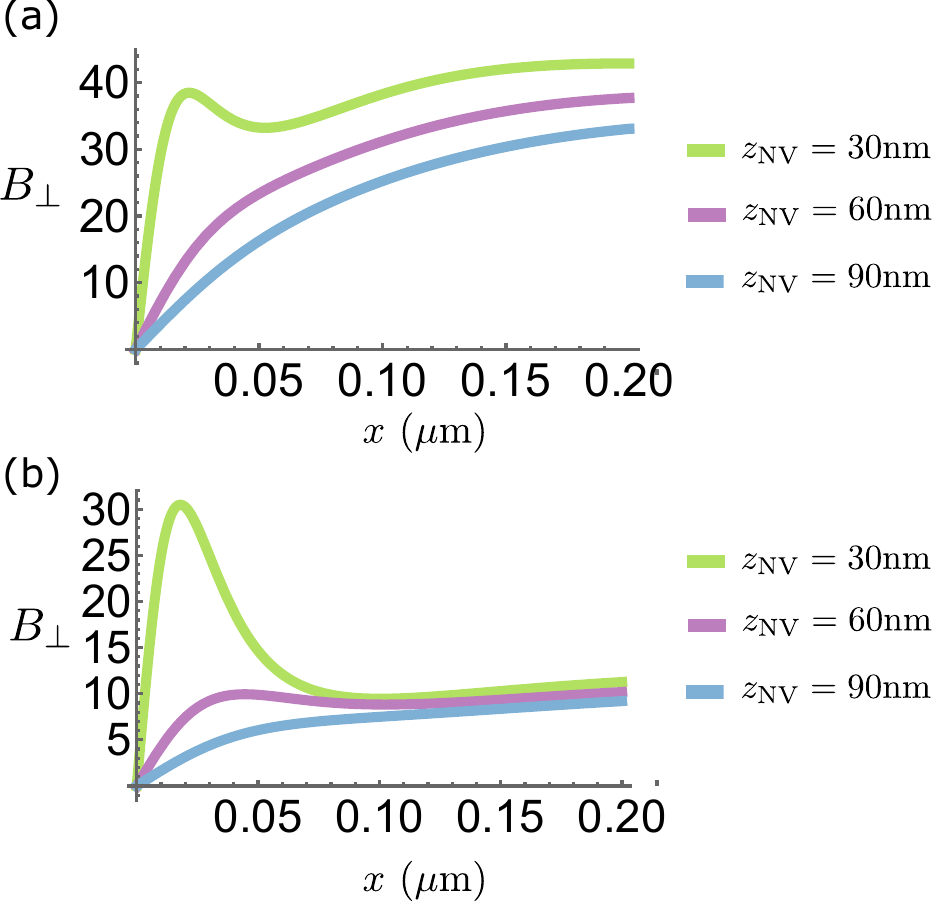}
    \caption{$B_{\perp}$ as a function of distance from the injector $x$ obtained using the no-slip boundary condition; (a) $w=100\text{nm}$ (b) $w=200\text{nm}$}
    \label{fig:fig5}
\end{figure}

 Now, we address the question whether the same behavior persists under no-slip boundary conditions. We plot in Fig.~\ref{fig:fig5}(a) $B_{\perp,z_{\text{NV}}}(x)$ computed with the no-slip boundary condition and the same parameters used for Fig.~\ref{fig:fig4} otherwise. One can see that the peak at small $x$ is less pronounced now. However, as shown n Fig.~\ref{fig:fig5}(b), if the film width is increased as, the peak is more manifest. These results suggest that the structure the we propose that viscosity introduces does survive in the no-slip boundary condition as well, but it may require a thicker magnetic insulator sample to observe.

  It is once again useful to build an intuition about the viscosity-induced peak in $B_{\perp}$ near the spin current injector as we did to understand $B_{\perp}(x)$ in the diffusive regime. As we observed in our discussion of the diffusive regime and the current profile shown in Fig.~\ref{fig:fig3}, the vertically injected current makes a turn near the injector, and the flow effectively becomes one-dimensional afterward. Meanwhile, viscosity can be understood as resistance to a change of fluid velocity. Hence, its effect is most manifest near the injector where the sharp change in the current direction occurs. The peak in $B_{\perp}$ can thus be understood as an extra spin potential gradient that the system generates to overcome the viscosity effect opposing the change in the current direction. A similar effect has been discussed in the context of charge transport \cite{Levitov2017}.

  \begin{figure*}
    \centering
    \includegraphics[width=0.75\linewidth]{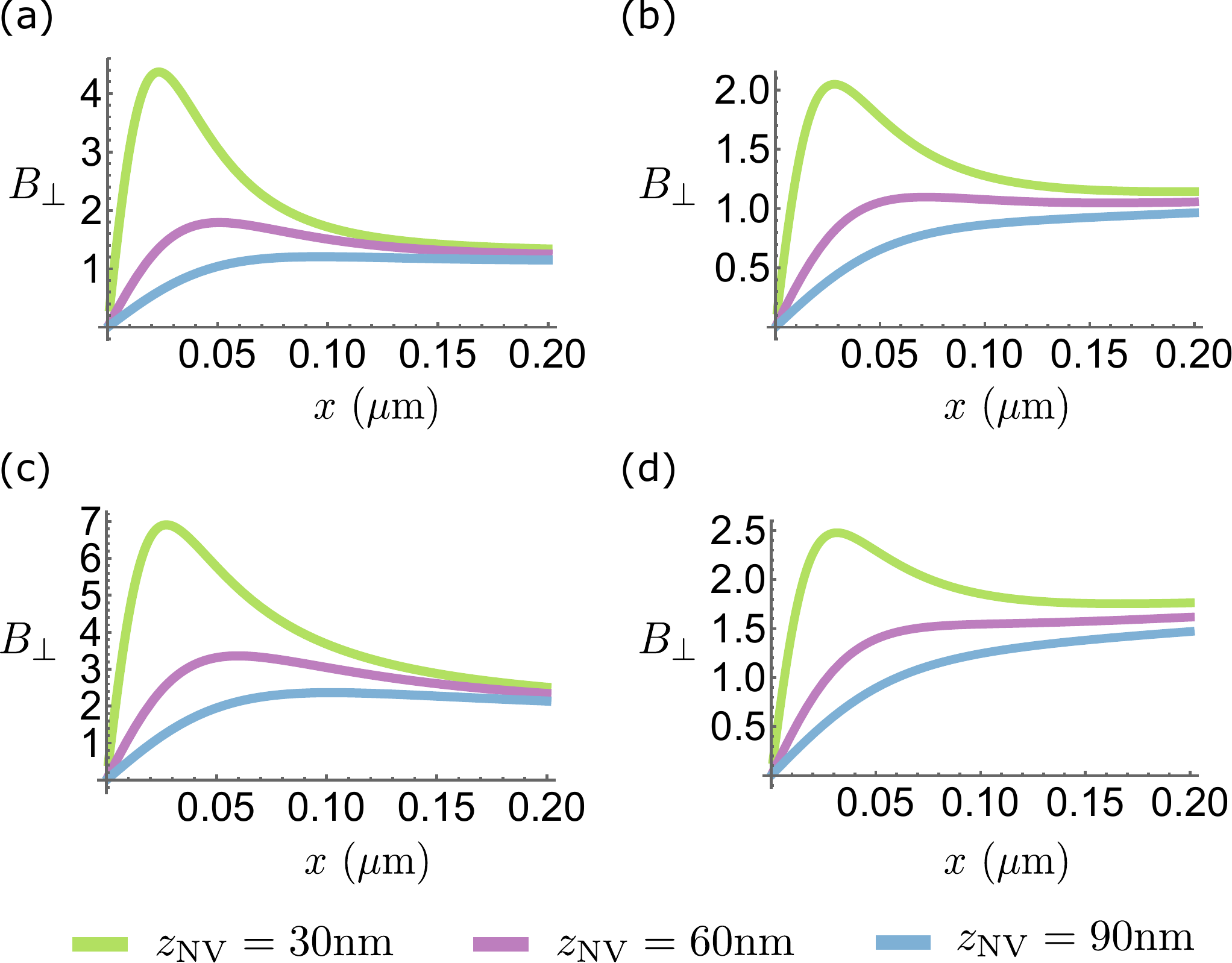}
    \caption{$B_{\perp,z_{\text{NV}}}(x)$ numerically obtained in the 2D hydrodynamic regime at (a) $(r_{\sigma}, r_{\eta}) = (0.025,0.025)$ (b) $(r_{\sigma}, r_{\eta}) = (0.4,0.025)$ (c) $(r_{\sigma}, r_{\eta}) = (0.025,0.4)$ (d) $(r_{\sigma}, r_{\eta}) = (0.4,0.4)$. The no-slip boundary condition, $l_{v} = 100\text{nm}$, $w=200\text{nm}$, $\xi_{s,\text{eff}} = 5\text{$\mu$m}$ are used for all plots.}
    \label{fig:fig6}
\end{figure*}

\subsection{2D hydrodynamic regime}

 The solution for the 2D hydrodynamics regime is given in Appendix~\ref{app:2Dhydro}. The solution, in addition to $l_{v}$, $w$, and $\xi_{s}$, also depends on $r_{\sigma}= \sigma_{s,\perp}/\sigma_{s,\text{eff}}$, which parameterizes the anisotropy in the effective spin conductivity between the out-of-plane and the in-plane directions, and $r_{\eta} = l_{v,\perp}/l_{v}$, which may be thought the anisotropy in momentum diffusion in the two different directions. Because the materials of interest are layered systems with weak interlayer coupling, we assume $r_{\sigma},r_{\eta}<1$ and explore the effect of $r_{\sigma}$ and $r_{\eta}$ for values ranging from $0.025$ to $0.4$. We find that $(i)$ the near-injector peak is present in the 2D hydrodynamic regime for all values of $(r_{\sigma}, r_{\eta})$ we explore $(ii)$ small $r_{\sigma}$ and large $r_{\eta}$ make the near-injector peak more pronounced. Fig.~\ref{fig:fig6} shows representative examples of $B_{\perp,z_{\text{NV}}}(x)$ for various anisotropy values.

\section{Conclusion}
\label{sec:Conc}

 In this paper, we developed a formalism to study current/potential distributions and the stray magnetic fields generated in viscous DC spin transport of magnetic insulators in the setup where spin current is vertically injected via the spin Hall effect of the metal interfaced with the magnet. Then, we applied the formalism to study how viscosity effect modifies the spatial profile of the stray magnetic field. Our analysis reveals that a sharp enhancement of the out-of-plane stray magnetic field near the current injector can serve as a diagnostics of viscous spin transport. As opposed to the broader peak in the magnetic field that arises due to finite spin diffusion lengths in the purely diffusive regime, the viscosity-induced feature depends sensitively on the vertical distance from the magnetic insulator.

 Our proposed protocol to detect the hydrodynamic effect from viscosity offers several advantages compared to the spin-noised based method proposed in Ref.~\onlinecite{spinhydro1} and \onlinecite{spinhydro2}. Our strategy is the most effective when the viscosity is large. However, the high viscosity strongly damps the sound modes and renders the spin noise-based method difficult. Additionally, the NV center relaxometry for probing spin noise has a poor $k$-space resolution when the magnetic insulator film thickness is comparable or greater than the distance between the NV-sample distance. Meanwhile, our strategy leverages the non-zero thickness of the film and is free from this issue. Hence, our protocol may be used to identify the hydrodynamic regimes of spins beyond those in atomically thin samples utilized in Ref.~\onlinecite{spinhydro2}. We expect that the method proposed in this paper will nicely complement the earlier noise-based approach.

 Having established that viscosity effect gives rise to an observable signature in spin transport, our work also poses various theoretical and experimental puzzles for future work. To list some of them:
  
  $(i)$ In charge transport, several studies have pointed out that some features of the current and potential profiles initially thought to be telltale signatures of viscous transport can also show up in the \textit{ballistic regime} \cite{Levitov2018, Cook2019}. The approach using Boltzmann equations to study ballistic-to-hydrodynamic crossover \cite{dejong1995,Superballistic, Levitov2018, Cook2019,Levitov2024} was crucial in shedding light on this issue in Fermi liquids. It would be insightful to develop a similar approach for magnonic or spinon-driven spin transport.

  $(ii)$ Here, we assumed that the simplest theory for the hydrodynamic regime in which only momentum and spin-$x$ are quasi-conserved variables. However, for example, in quantum spin liquids, various symmetries that emerge in the IR description (most notably, U$(1)$ monopole conservation) may lead to additional hydrodynamic modes and dissipative coefficients \cite{spinhydro3,Banerjee2025}. In a similar vein, assuming a weaker point-group symmetiries instead of full rotation symmetry introduces extra viscosity terms in the hydrodynamic equations \cite{Link2018,Cook2019,Rao2020,Varnavides2020}. Systematically investigating how these additional features modify the DC spin transport signatures would be an important step forward as well.

  $(iii)$ For the experimental guidance, it would be valuable to develop a method to systematically compute the strength of the viscosity of the magnetic insulators.

\acknowledgements 

 We acknowledge funding from AFOSR MURI grant no. FA9550-21-1-0429 to Eun-Ah Kim. This research is partly funded by the Gordon and Betty Moore Foundation's EPiQS Initiative, Grant GBMF10436 to  Eun-Ah Kim. We acknowledge helpful discussions with Aaron Hui, Katja Nowack, and S. Raghu, and Eun-Ah Kim.

\bibliography{ref}

\clearpage
\onecolumngrid

\appendix
\section{Remarks on the time-reversal symmetry in spin hydrodynamics}
\label{app:timereversal}
 In this appendix, we comment on why $\mathcal{T}$-breaking is necessary for our consideration of the hydrodynamic regime and motivate  $\mathcal{CT}$ symmetry we invoked at the main text based on that reasoning.

 In the hydrodynamic regime of spin transport we consider, one expects that momentum $\mathbf{P}$ is a long-lived hydrodynamic variable and that the spin current $\mathbf{j}_{s_{x}}$ is primarily carried by the momentum. The second statement essentially means that in the regime of our interest, we expect the constitutive relation for $\mathbf{j}_{s_{x}}$ to contain a term linear in $\mathbf{P}$. This statement is particularly important for DC spin transport to be subject to a strong viscosity effect.
 
 However, $\mathbf{P}$ is odd under time reversal symmetry $\mathcal{T}$, while $\mathbf{j}_{s_{x}}$ is even under $\mathcal{T}$. One can readily see that \textit{to observe the viscous effect in spin transport, time-reversal symmetry must be broken, either spontaneously or explicitly}. In the magnon-mediated spin transport for ordered magnets, $\mathcal{T}$ is spontaneously broken, and this criterion is fulfilled somewhat trivially. However, for the purpose of probing the spin liquid physics, one must always assume the presence of an external field $B_{x}$ so that $(i)$ the full spin rotation symmetry $\text{SU}(2)$, if present, is reduced to $\text{U}(1)$ and $(ii)$ $\mathcal{T}$ is explicitly broken. 
  
 While we set $\mathcal{T}$ to be spontaneously or explicitly broken in our setup, it would be still desirable to constrain the theory with a time-reversal-like symmetry. The minimal way of achieving this goal is to assume the presence of spin-flip symmetry $\mathcal{C}$ $S_{x} \rightarrow -S_{x}$ in the underlying Hamiltonian (without external magnetic field) and take $\mathcal{CT}$ as an effective time-reversal symmetry valid in the infra-red (IR) ($\mathcal{CT}$ is taken to be valid even after applying the external magnetic field $B_{x}$). Note that that $\mathbf{j}_{s_{x}}$ is odd under $\mathcal{CT}$ due to the spin-flip operation and now compatible with a linear-in-$\mathbf{P}$ term in the constitutive relation.

 The question that remains is whether $\mathcal{C}$-symmetry in the underlying Hamiltonian and $\mathcal{CT}$ in the infra-red physics are physical. As partial answers to this question, we note the following: $(i)$ If the system has the full $\text{SU}(2)$ spin rotation symmetry, a $\pi$ rotation of spins around the $y$-axis or the $z$-axis naturally implements $\mathcal{C}$. $(ii)$ for ferromagnets and antiferromagnets whose spontaneous magnetizations are parallel to the $x$-axis, $\mathcal{CT}$ map the ordering vector to itself and is a valid IR symmetry. $(iii)$ Similarly, since the operator $S_{x}$ is even under $\mathcal{CT}$, applying the in-plane field $B_{x}$, equivalent to adding a term $B_{x}S_{x}$ to the Hamiltonian, breaks $\mathcal{C}$ and $\mathcal{T}$ individually but not $\mathcal{CT}$. Hence, we find $\mathcal{C}$-symmetry and IR $\mathcal{CT}$ symmetry to be compatible with the physical scenarios we consider and take them to be valid symmetries to assume.

\section{No-slip solution in the 3D hydrodynamic regime}
\label{app:noslip}

 Using the no-slip boundary condition to compute $A(k)$, $B(k)$, $C(k)$, $D(k)$ in Eq.~\eqref{eq:3Dgeneral} instead of the no-stress boundary condition employed in the main text, one can obtian the following spin chemical potential distribution, where $q_{s} = \sqrt{k^{2} + \xi_{s,\text{eff}}^{-2}}$ and $q_{v} = \sqrt{k^{2} + l_{v}^{-2}}$:
\begin{equation}
\tilde{\mu}_{s_{x},k}(z) =  I_{s} (1 - l_{v}^{2} \xi_{s,\text{eff}}^{-2})  \frac{ \left[ \frac{q_{s} q_{v} }{k^{2}} \sinh q_{v}  w  - \sinh q_{s} w\right] \cosh q_{s} z + \left[ \cosh q_{s}  w  - \cosh q_{v} w \right] \sinh q_{s} z  }{q_{s} \left[ \left( \frac{q_{v}q_{s}}{k^{2}} +\frac{k^{2}}{q_{v}q_{s}}   \right) \sinh q_{v} w \sinh q_{s} w + 2 - 2 \cosh q_{v} w \cosh q_{s} w\right]}. 
\end{equation}
Plugging this solution into Eq.~\eqref{eq:Bperpk} and integrating over $z'$, one can obtain the stray magnetic field profile:
\begin{equation}
\begin{split}
& B_{\perp, z_{\text{NV}}}(x) =  -\frac{\mu_{0} I_{s} C_{s} }{\pi \sigma_{s,\text{eff}}}  \left( 1 - l_{v}^{2} \xi_{s,\text{eff}}^{-2} \right) \int_{0}^{\infty} dk \,   \frac{ k \sin kx  e^{-k z_{\text{NV}}}}{N(k)} \bigg[  \frac{q_{s} q_{v}}{4}  (1 - e^{-2 q_{v} w}) \left( \frac{1 - e^{-(k+q_{s}) w}}{k + q_{s}}  + \frac{e^{- 2 q_{s} w} - e^{-(k+q_{s}) w}}{k - q_{s}}  \right) \\
& - \frac{k^{2}}{4} (1 + e^{-2 q_{v} w})  \left( \frac{1 - e^{-(k+q_{s}) w}}{k + q_{s}}  - \frac{e^{- 2 q_{s} w} - e^{-(k+q_{s}) w}}{k - q_{s}}  \right)  + \frac{k^2}{2} e^{- (q_{v} + q_{s})  w}  \frac{1 - e^{-(k+q_{s}) w}}{k + q_{s}}  - \frac{k^2}{2} e^{- q_{v} w}  \frac{e^{-q_{s} w} - e^{- k w} }{k - q_{s}}  \bigg],
\end{split} 
\end{equation}
where we define $N(k)$ as the following:
\begin{equation}
N(k) = q_{s} \left[  \frac{1}{4}\left( q_{v} q_{s} + \frac{k^{4}}{q_{v} q_{s}}\right) (1 - e^{-2 q_{v} w}) (1 - e^{-2 q_{s} w})  + 2 k^{2} e^{-(q_{v}+q_{s})w} - \frac{1}{2} k^{2} (1 + e^{-2 q_{v} w}) (1 + e^{-2 q_{s} w})  \right].
\end{equation}

\section{Solutions to the boundary value problems for the 2D hydrodynamic regime}
\label{app:2Dhydro}

\subsection{Solutions to the differential equations}

 Here, we briefly look at how to solve the differential equations through separation of variable methods. First, it is convenient to introduce the following rescaled variables: 
 \begin{equation}
\tilde{x} = x/l_{v\parallel}, \quad \tilde{z} = z/l_{v\perp}, \quad \mu_{r} = \mu_{s_{x}}\sigma_{s,\perp}/l_{v\perp}^{2}, \quad j_{r} = j_{s_{x},x}/l_{v\parallel}.
\end{equation}
Also, introduce the following dimensionless parameters:
\begin{equation}
\tilde{\alpha} = \frac{C_{s} l_{v\perp}^{2}}{\sigma_{s,\perp}\tau_{s}} =\frac{l_{v}^{2} r_{\eta}^{2}}{ r_{\sigma} \xi_{s,\text{eff}}^{2}} , \quad R = \frac{\sigma_{s,\text{eff}} l_{v\perp}^{2}}{\sigma_{s,\perp} l_{v}^{2}} = r_{v}^{2}/r_{\sigma} ,\quad \tilde{w} = w/l_{v\perp}.
\end{equation}
In the definition of $\tilde{\alpha}$, generalizing the definition of the spin diffusion length given at the beginning of Sec.~\ref{sec:solutiondiffusive} for the diffusive case, we recognized $\sqrt{\sigma_{s,\parallel} \tau_{s}C_{s}^{-1}}$ as the effective spin diffusion length $\xi_{s,\text{eff}}^{2}$ along the $xy$-plane in this setup.

 In terms of the rescaled variables and dimensionless parameters introduced above, our differential equations become:
\begin{equation}
-\partial_{\tilde{z}}^{2} \mu_{r} + \partial_{\tilde{x}} j_{r} = - \tilde{\alpha} \mu_{r} , \quad - j_{r} = R \partial_{\tilde{x}} \mu_{r} - \nabla^{2} j_{r}.
\end{equation}
The most convenient way to solve this system of partial differential equations is to take the partial derivative $\partial_{\tilde{x}}$ to the both sides of the first equation, and replace $\mu_{r}$ with expressions involving $j_{r}$ only using the second equation. This procedure gives the following fourth-order differential equation for $j_{r}$:
\begin{equation}
\partial_{\tilde{z}}^{4} j_{r} - (\tilde{\alpha} + 1 - \partial_{\tilde{x}}^{2}) \partial_{\tilde{z}}^{2} j_{r} - (R \partial_{\tilde{x}}^{2} + \tilde{\alpha}  \partial_{\tilde{x}}^{2} - \tilde{\alpha}) j_{r}= 0.
\end{equation}
 As we did before, we can solve this differential equation by introducing Fourier transform of $j_{r}$,  $\tilde{j}_{r,k}(\tilde{z}) = \int_{-\infty}^{\infty} d\tilde{x} \, e^{i k \tilde{x}}  j_{r}(\tilde{x},\tilde{z})$. To present the solution, we first introduce:
\begin{equation}
\begin{split}
& \lambda_{1}(k) = \sqrt{\frac{k^2 + 1 + \tilde{\alpha} }{2} + \sqrt{\left( \frac{k^2 + 1 + \tilde{\alpha} }{2}\right)^{2} - (R + \tilde{\alpha})k^{2} - \tilde{\alpha} }}, \\
& \lambda_{2}(k) = \frac{ \sqrt{ (R + \tilde{\alpha})k^{2} + \tilde{\alpha}} }{ \lambda_{1}(k) }.
\end{split}
\end{equation}
The general solution to the equation can be written as:
\begin{equation}
\tilde{j}_{r,k}(\tilde{z}) = A(k) \cosh \lambda_{1}(k) \tilde{z} + B(k) \sinh \lambda_{1}(k) \tilde{z} + C(k) \cosh \lambda_{2}(k) \tilde{z} + D(k) \sinh \lambda_{2}(k) \tilde{z}.
\end{equation}
$A(k)$, $B(k)$, $C(k)$, $D(k)$ can be determined by matching the boundary conditions at $z=0,-w$.

 By plugging in the above solution for $\tilde{j}_{r,k}(\tilde{z})$ to the original system of differential equations, one can show that $\tilde{\mu}_{r,k}(\tilde{z}) =  \int_{-\infty}^{\infty} d\tilde{x} \, e^{i k \tilde{x}}  \tilde{\mu}_{r}(\tilde{x},\tilde{z})$ takes the following form:
\begin{equation}
\tilde{\mu}_{r,k}(\tilde{z}) = \frac{k^{2} + 1 - \lambda_{1}^{2}(k)}{i k R}\left[ A(k) \cosh \lambda_{1}(k) \tilde{z} + B(k) \sinh \lambda_{1}(k) \tilde{z} \right] +  \frac{k^{2} + 1 - \lambda_{2}^{2}(k)}{i k R} \left[ C(k) \cosh \lambda_{2}(k) \tilde{z} + D(k) \sinh \lambda_{2}(k) \tilde{z} \right].
\end{equation}
In the below, we present the explicit form of the solution for $\tilde{\mu}_{r,k}(\tilde{z})$ and $B_{\perp, z_{\text{NV}}}(x)$ computed from the solution for the no-stress boundary condition and no-slip boundary condition.
\subsection{No-stress}

 By imposing the no-stress boundary condition, one can obtain the following spin pontential profile solution:
\begin{equation}
\begin{split}
\tilde{\mu}_{r,k}(\tilde{z}) & = \frac{ I_{s}}{ l_{v\parallel}l_{v\perp}}\frac{k^{2} + 1 -\lambda_{1}(k)^{2} }{\lambda_{2}(k)^{2} - \lambda_{1}^{2}(k)}  \frac{\cosh \lambda_{1}(k) \tilde{z}}{\lambda_{1}(k) \sinh \lambda_{1}(k) \tilde{w}} + (1 \leftrightarrow 2).
\end{split}
\end{equation}
 The symbol $(1 \leftrightarrow 2)$ means the term obtained from the previous one by exchanging the role of $\lambda_{1}(k)$ and $\lambda_{2}(k)$. Inserting this solution to Eq.~\eqref{eq:Bperpk} (be aware that $\tilde{\mu}_{r,k}(\tilde{z}) $ is Fourier-transformed with respect to the rescaled coordinates. One should convert back to the original coordinates to apply this formula) and integrating over $z'$, one obtains: 
\begin{equation}
\begin{split}
   B_{\perp,z_{\text{NV}}}(x) & = -\frac{\mu_{0} I_{s} C_{s} R }{\pi \sigma_{s,\text{eff}}}   \int_{0}^{\infty}  dk \,   \frac{e^{- k z_{\text{NV}} / l_{v}} k \sin \frac{k x}{l_{v}} }{\lambda_{1}(k) ( 1 - e^{-2 \lambda_{1}(k) \tilde{w}})}   \frac{k^{2} + 1 - \lambda_{1}^{2}(k) }{\lambda_{2}(k)^{2} - \lambda_{1}^{2}(k)} \left[ \frac{ 1- e^{-(r_{\sigma} k+\lambda_{1}(k))\tilde{w} }}{r_{\sigma} k+\lambda_{1}(k)} + \frac{e^{ - 2 \lambda_{1}(k) \tilde{w} } - e^{-(k+\lambda_{1}(k))\tilde{w}} }{r_{\sigma} k-\lambda_{1}(k) } \right]   \\
  & \quad \quad   + (\lambda_{1}(k) \leftrightarrow \lambda_{2}(k) ).
\end{split}
 \end{equation}

\subsection{No slip}

 For the better visibility of the equations we present here, we define the following functions:
\begin{equation}
\begin{split}
& Q_{1}(k) = \lambda_{1}(k)(k^{2} + 1 - \lambda_{1}^{2}(k)), \quad Q_{2}(k) = \lambda_{2}(k)(k^{2} + 1 - \lambda_{2}^{2}(k)) \\
& N(k) =  \left( \frac{Q_{1}(k)}{Q_{2}(k)} + \frac{Q_{2}(k)}{Q_{1}(k)} \right)\sinh \lambda_{1}(k) \tilde{w}  \sinh \lambda_{2}(k) \tilde{w} + 2 - 2 \cosh \lambda_{1}(k) \tilde{w}  \cosh \lambda_{2}(k) \tilde{w}.
\end{split}
\end{equation}

 Utilizing the no-slip boundary conditions gives the following spin chemical potential:
\begin{equation}
\begin{split}
\tilde{\mu}_{r,k}(\tilde{z}) = \frac{I_{s}}{ l_{v\parallel} l_{v\perp}}  \frac{1}{N(k)} & \Bigg[ \left(\frac{Q_{1}(k) }{Q_{2}(k)} \sinh \lambda_{2}(k) \tilde{w}- \sinh \lambda_{1}(k) \tilde{w} \right) \lambda_{1}(k) ^{-1} \cosh \lambda_{1}(k)z \\
& \quad \quad \quad \quad + \left(\cosh \lambda_{1}(k) \tilde{w} - \cosh \lambda_{2}(k) \tilde{w} \right)\lambda_{1}(k) ^{-1} \sinh \lambda_{1}(k) z   \Bigg] +  (1 \leftrightarrow 2).
 \end{split} 
\end{equation}
$(1 \leftrightarrow 2)$ means the term obtained from the previous term by exchanging the role of $(\lambda_{1}(k), Q_{1}(k))$ and $(\lambda_{2}(k), Q_{2}(k))$. Through our usual procedure, one can also obtain $B_{\perp,z_{\text{NV}}}(x)$ from the above solution, which is evaluated to be:
\begin{equation}
\begin{split}
   B_{\perp,z_{\text{NV}}}(x) 
  & = -\frac{\mu_{0} I_{s} C_{s} R }{\pi \sigma_{s,\text{eff}} }   \int_{0}^{\infty}  dk \, \frac{ e^{- k z_{\text{NV}} / l_{v}} k \sin \frac{k x}{l_{v}} }{  e^{-(\lambda_{1}(k) + \lambda_{2}(k)) \tilde{w} }N(k)} \\
  & \quad \quad \quad \quad \quad \quad \quad \quad \quad \quad  \bigg[ \frac{Q_{1}(k) (1 - e^{-2 \lambda_{2}(k) \tilde{w}}) }{4 \lambda_{1}(k) Q_{2}(k)}  \left( \frac{1 - e^{- (\lambda_{1}(k)  + r_{\sigma}k ) \tilde{w} } }{\lambda_{1}(k)  + r_{\sigma}k } + \frac{  e^{ - 2 \lambda_{1}(k) \tilde{w}} - e^{- (\lambda_{1}(k)  + r_{\sigma}k ) \tilde{w} }  }{-\lambda_{1}(k)  + r_{\sigma}k } \right) \\
 & \quad \quad \quad \quad \quad \quad \quad \quad \quad \quad - \frac{(1 + e^{-2 \lambda_{2}(k) \tilde{w}}) }{4 \lambda_{1}(k) }  \left( \frac{1 - e^{- (\lambda_{1}(k)  + r_{\sigma}k ) \tilde{w} } }{\lambda_{1}(k)  + r_{\sigma}k } - \frac{  e^{ - 2 \lambda_{1}(k) \tilde{w}} - e^{- (\lambda_{1}(k)  + r_{\sigma}k ) \tilde{w} }  }{-\lambda_{1}(k)  + r_{\sigma}k } \right) \\
 &  \quad \quad \quad \quad \quad \quad \quad \quad \quad \quad   + \frac{e^{- \lambda_{2}(k)\tilde{w} } }{2 \lambda_{1}(k) }  \left( \frac{e^{- \lambda_{1}(k)\tilde{w} } - e^{- (2 \lambda_{1}(k)  + r_{\sigma}k ) \tilde{w} } }{\lambda_{1}(k)  + r_{\sigma} k } - \frac{  e^{- \lambda_{1}(k)\tilde{w} } - e^{- r_{\sigma} k  \tilde{w} }   }{-\lambda_{1}(k)  + r_{\sigma} k } \right)  \bigg] \\
 & \quad \quad \quad \quad \quad \quad \quad \quad \quad \quad + (1 \leftrightarrow 2).
\end{split}
 \end{equation}

\end{document}